\documentclass[twocolumn,english,aps,prx,floatfix,amssymb,superscriptaddress,longbibliography]{revtex4}
\usepackage[latin9]{inputenc}
\usepackage{verbatim}
\usepackage{float}
\usepackage{amsmath}
\usepackage{amssymb}
\usepackage{graphicx}
\usepackage{color}
\usepackage{xcolor}
\usepackage{tikz}
\usepackage[normalem]{ulem}
\newcommand{\jp}[1]{\textcolor{blue}{#1}}

\newcommand{\ket}[1]{\ensuremath{\left| #1 \right>}}

\usepackage{bbold}

\usepackage[bookmarks=false,linkcolor=blue,urlcolor=blue,colorlinks,citecolor=blue]{hyperref}
\makeatletter

\newcommand{\be}{\begin{equation}}
\newcommand{\ee}{\end{equation}}
\newcommand{\bea}{\begin{eqnarray}}
\newcommand{\eea}{\end{eqnarray}}

\begin{document}

\title{Spin fractionalization and zero modes in the spin-$\frac{1}{2}$ XXZ chain with boundary fields}
\author{Parameshwar R. Pasnoori}
\affiliation{Department of Physics, University of Maryland, College Park, MD 20742, United
States of America}
\affiliation{Laboratory for Physical Sciences, 8050 Greenmead Dr, College Park, MD 20740,
United States of America}

\author{Yicheng Tang}
\affiliation{Department of Physics and Astronomy, Center for Materials Theory, Rutgers University,
Piscataway, NJ 08854, United States of America}

\author{Junhyun Lee}
\affiliation{Department of Physics and Astronomy, Center for Materials Theory, Rutgers University,
Piscataway, NJ 08854, United States of America}

\author{J. H. Pixley}
\affiliation{Department of Physics and Astronomy, Center for Materials Theory, Rutgers University,
Piscataway, NJ 08854, United States of America}
\affiliation{Center for Computational Quantum Physics, Flatiron Institute, 162 5th Avenue, New York, NY 10010, USA}

\author{Natan Andrei}
\affiliation{Department of Physics and Astronomy, Center for Materials Theory, Rutgers University,
Piscataway, NJ 08854, United States of America}
\email{pradip.kattel@rutgers.edu}
\email{pparmesh@umd.edu}

\author{Patrick Azaria}
\affiliation{Laboratoire de Physique Th\'orique de la Mati\`ere Condens\'ee, Sorbonne Universit\'e and CNRS, 4 Place Jussieu, 75252 Paris, France}
\date{\today}

\begin{abstract}
In this work we argue that the antiferromagnetic  spin $\frac{1}{2}$ XXZ chain in the gapped phase with boundary magnetic fields  hosts fractional spin $\frac{1}{4}$ at its edges. 
Using a combination of Bethe ansatz and the density matrix renormalization group we show that these fractional spins are sharp  quantum observables in both the ground and the first excited state as the associated fractional spin operators have zero variance. 
In the limit of zero edge fields, we argue that these fractional spin operators once projected onto the low energy subspace spanned  by the ground state and the first excited state, identify with the strong zero energy mode discovered by P. Fendley \cite{Fendley}.

\end{abstract}
\maketitle
\section{Introduction}

Since the discovery of solitons carrying half of the electron charge \cite{JackiwRebbi,ssh} it has been widely
recognized \cite{goldstonewilcezk,kivelsonschriefer,Jackiw} that some states of matter can be characterized by fractional quantum numbers. 
Maybe the most celebrated example is the fractional quantum Hall state where quasiparticles carry fractional charges \cite{Laughlin,Stormer}. 
Other prominent examples coming from topological phases with short range topological order, such as symmetry protected
topological (SPT) systems in one dimension, include spin-$1/2$ edge states in the spin one Haldane chain \cite{HALDANE,AKLT} as well as 
spin-$1/4$ zero energy 
modes (ZEM) localized at the edges of one dimensional spin triplet superconductors \cite{Keselman2015,PAA1,PAA2}. In higher dimensions, surface states in topological insulators
as well as disordered magnetic systems like spin ice \cite{spinicefrac} and spin liquids also exhibits signatures of fractionalization \cite{spinliquidfrac}.

In this work we shall demonstrate that fractionalization can also occur in more conventional magnetic systems which exhibit long range 
magnetic order. To this end we shall consider the paradigmatic
XXZ spin $1/2$ chain with magnetic fields at its two edges, that we solve exactly with the Bethe ansatz and numerically using the density matrix renormalization group (DMRG),
and show that in the low energy sector it hosts  quantum spin-$1/4$
states localized at the  edges. We shall further  argue that this fractional quarter spins are sharp quantum observables. We believe that this  result might have some impact  in understanding dynamics \cite{Wensho,Mitra,Pozsgay,Mesty,Yuzbashyan,Santos,Mallick},
heat and spin transport \cite{Bertini,Collura,Moore}. 

\begin{center}
\begin{figure}[h!]
    \includegraphics[width=0.45\textwidth]{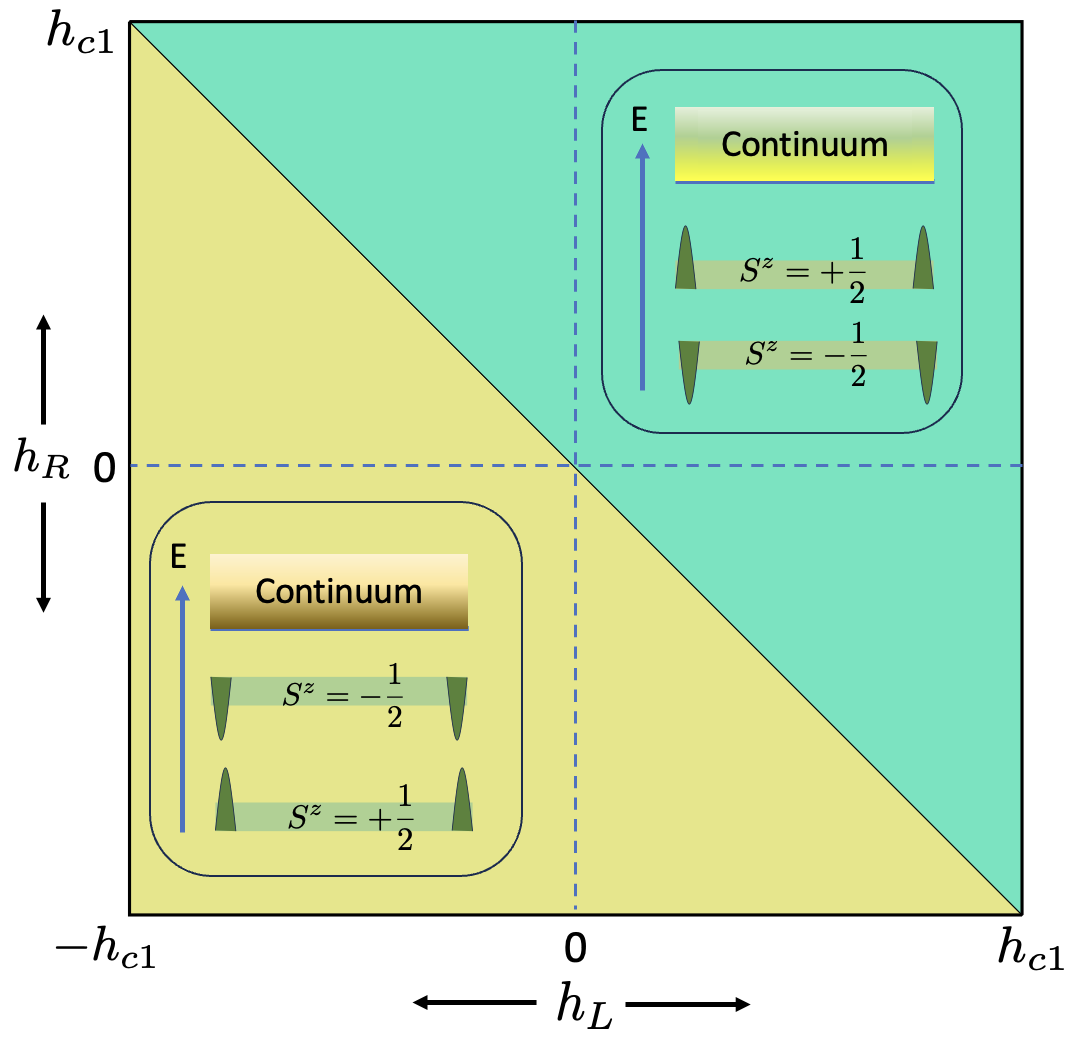}
    \caption{Ground state phase diagram of the XXZ model with edge fields smaller than the critical field $|h_{L,R}| < h_c=\Delta-1$  and for an odd number of sites. In each of the two phases separated by the line $h_L+h_R=0$ we show the ground state as well as the first  excited state which is a midgap state  below the continuum. Both states host fractional quarter spins ${\cal S}_{L,R}=\pm 1/4$ at both edges of the chain. These fractional spins  are sharp quantum observables which reconstruct the total spin of each state $S^z={\cal S}_{L}+{\cal S}_{R} =\pm 1/2$. The $\pm 1/4$ quarter spins are depicted  by triangles pointing  upward and downward respectively. On the separatrix $h_L+h_R=0$ there is  spontaneous symmetry breaking and the edge spin operator becomes a zero energy mode. 
    }
    \label{pdodd}
\end{figure}
\end{center}

We consider the XXZ hamiltonian with boundary magnetic fields $(h_{L}, h_{R})$ at the left and the right  edges of an open chain
\bea
\label{XXZ}
H&=&\sum_{j=1}^{N-1} \sigma^x_j\sigma^x_{j+1}+\sigma^y_j\sigma^y_{j+1}+\Delta (\sigma^z_j\sigma^z_{j+1}-1)\nonumber\\&+&h_L\sigma^z_1+h_R\sigma^z_N
\eea
where $\sigma_j^{x,y,z}$ are the Pauli matrices and $\Delta > 1$ is the anisotropy parameter. In the limit where the boundary fields are zero,  on top of being $U(1)$ symmetric, (\ref{XXZ}) is space parity  $\mathbb{P}$ and time reversal $\mathbb{T}$ invariant. It is also invariant under the 
 $\mathbb{Z}_2=\{1,\tau \}$ spin flip symmetry, i.e:
 $[H, \tau]=0$ where $\tau=\Pi_1^N \sigma_j^x$. For generic non zero boundary fields $h_{L, R},\neq 0$, both  $\mathbb{P}$ and $\mathbb{Z}_2$ symmetries are explicitly broken. However,  on the two lines $h_{L}= \pm  h_{ R}$ the  hamiltonian (\ref{XXZ})   displays $\mathbb{P}$ and $\mathbb{P} \circ \mathbb{Z}_2 $ symmetries respectively.

The Hamiltonian in Eq.~(\ref{XXZ}) is integrable by the method of the Bethe ansatz for arbitrary boundary fields
$h_{ L,  R}$ and $\Delta$, which is used in the present paper to determine the low energy eigenstates analytically. The system with periodic boundary conditions was first solved by Bethe \cite{Bethe1931} in the isotropic limit, $\Delta\rightarrow 1$. The solution was later extended to include anisotropy along the z-direction \cite{Orbach,Walker,YangYang1,YangYang2,YangYang3,BABELON}.  In the gapped regime \jp{($\Delta>1$)} it exhibits a continuous $U(1)$ symmetry and also a discrete $\mathbb{Z}_2$ spin flip symmetry. The discrete $\mathbb{Z}_2$ symmetry is spontaneously broken \cite{Olav} and in the thermodynamic limit the system exhibits two degenerate symmetry broken ground states \cite{Takahashi}. The Bethe ansatz method to include the boundaries was developed in \cite{Alcaraz,Cherednik,Sklyanin} and the ground state and boundary excitations in various bulk phases exhibited by the XXZ spin chain were found in \cite{skorik,kapustinxxz,xxzbound2019,Nassar_1998}. An independent method to diagonalize the Hamiltonian using vertex operators was developed in \cite{Davies,JIMBO}, and was later extended to include the boundary fields in \cite{Miwa} where the boundary S-matrix and the integral formula for correlation functions have been found. Recently new band structures in the spectrum at large anisotropies have been found \cite{sharma}. 

Numerically we solve the model using DMRG,  implemented through the TeNPy software~\cite{tenpy}, that allows access to the ground state and midgap state with arbitrary precision thanks to the gapped nature of both of these states.  We take a maximum bond dimension of 400, with a minimal singular value decomposition cut off of $10^{-10}$, and converge the energy up to a maximal energy error on the order of $\sim10^{-10}$.

When $\Delta > 1$ the ground state
$|g\rangle$ displays antiferromagnetic order with non zero  staggered magnetization $\sigma= \lim_{N\rightarrow \infty}N^{-1} \sum_{j=1}^N (-1)^j \langle g|\sigma_j^z |g\rangle$ and is gapped.
Indeed, for all values of the edge fields, there is a gap ($m$) in the spectrum  to single particle spin $1/2$ spinon excitations
\be
m=\sinh \gamma \sum_{n \in \mathbb{Z}} \frac{(-1)^n}{\cosh \gamma n},\; \Delta=\cosh{\gamma}.
\ee
However at low fields, i.e: $|h_{ L,  R}| < \Delta -1$, the lowest excited state is a {\it midgap} state $|e\rangle$ which lies below the continuum.    We obtain the bound state energies \footnote{Note that a different expression was obtained in \cite{kapustinxxz,xxzbound2019}.} which are given by
\bea m_{\alpha}=h_{\alpha}+ \sinh\gamma \sum_{n \in \mathbb{Z}}(-1)^{n}\; \frac{\sinh(\gamma {\epsilon}_{\alpha}|n|)}{\cosh \gamma n}e^{-\gamma|n|}
\label{bsenergy}\eea
where $\alpha=(L,R)$. 
This midgap state is reminiscent of the existence of spin $1/2$ boundary bound states, localized at the two left and right edges.
The spin quantum numbers and energies of the ground state as well as the midgap state 
depend on both the parity of the number of sites, $(-1)^N$, as well as on the boundary fields $h_{L,R}$. When $N$ is odd, the two states $|g\rangle$ and $|e\rangle$ have opposite total spins $S^z=\pm 1/2$. Taking as a reference state the $|-\frac{1}{2}\rangle$ state with energy $E_0$, the $|+\frac{1}{2}\rangle$ state is obtained by adding a localized  bound state at each {\it edge}. This state has energy $E_0+ m_L+m_R$. Depending on the edge magnetic fields, and hence on the sign of $m_L+m_R$, the ground state  and the midgap state ($|g\rangle$,$|e\rangle$)
are ($|-\frac{1}{2}\rangle,  |+\frac{1}{2}\rangle$) when $h_L+ h_R > 0$ and
($|\frac{1}{2}\rangle,  |-\frac{1}{2}\rangle$) when $h_L+ h_R < 0$. Notice that
on the line   $h_L+ h_R = 0$, the two states $|\pm \frac{1}{2}\rangle$ are degenerate. In the $N\rightarrow \infty$ limit,  there is spontaneous symmetry breaking (SSB) of the $\mathbb{P} \circ \mathbb{Z}_2 $ symmetry. In the particular case of zero edge fields, both $\mathbb{P} $ and $ \mathbb{Z}_2 $ symmetries are spontaneously broken.
For $N$ even  both ($|g\rangle$,$|e\rangle$) states have total spins $S^z= 0$ and the bound state construction is presented in the Appendix. We display in Fig.(\ref{pdodd}) the  phase diagram for low  fields for an odd number of sites $N$.

\section{Spin Profiles}
\begin{figure}
    \centering
    \includegraphics[width=0.47\textwidth]{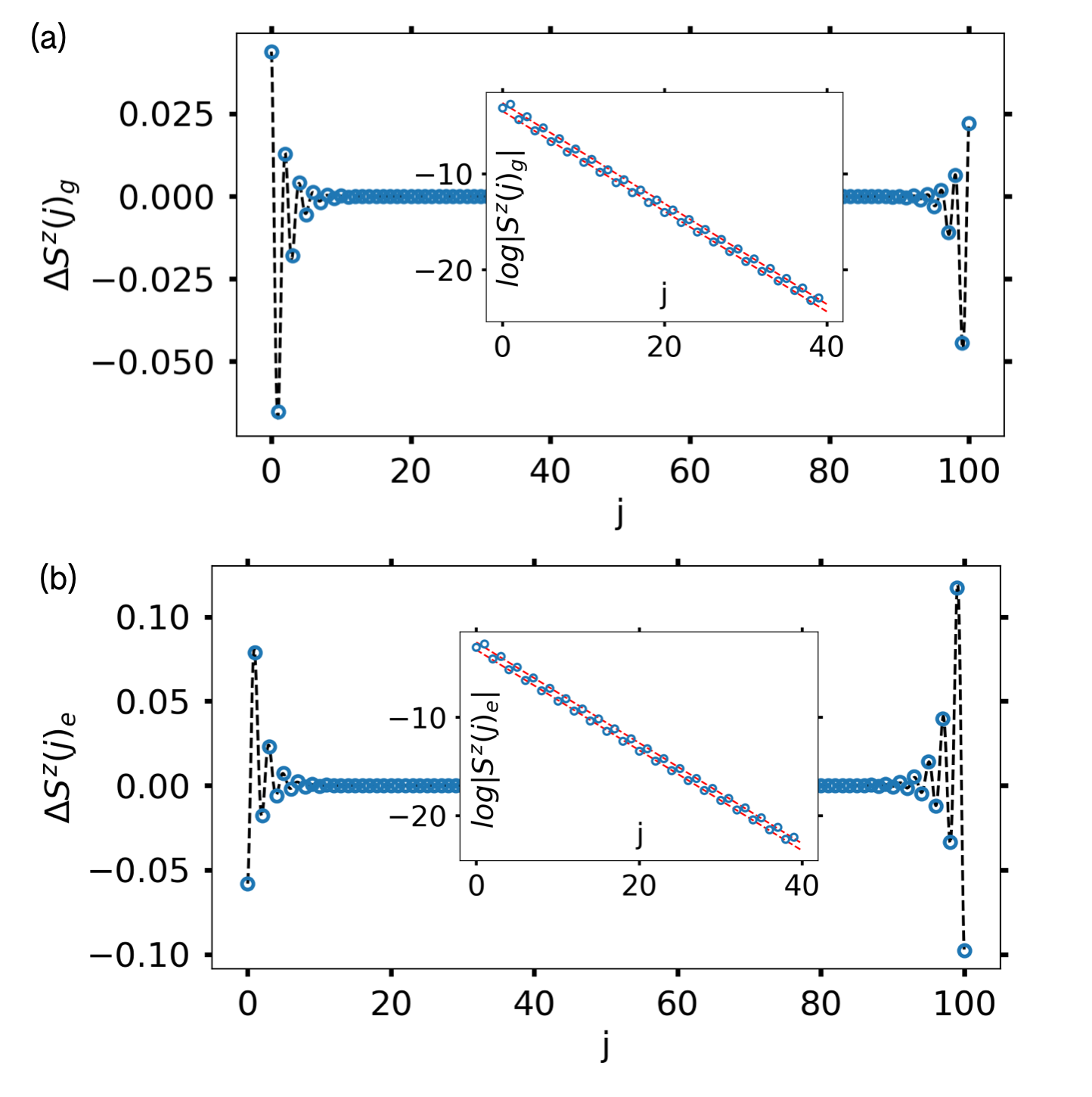}
    \caption{Spin profile $\Delta S^z(j)$ and the fitting of ansatz (a) in the ground  state $|g\rangle$ with total spin $S^z=-\frac{1}{2}$  (b) in the midgap  state $|e\rangle$ with total spin $S^z=\frac{1}{2}$ for model parameters $N=101$, $h_L=0.1,h_R=0.5$ and $\Delta=3$. The insets show that the relative spin are indeed localized exponentially on the edge, with red dashed lines representing excellent linear fits on the log-scale.}
    \label{DS_odd}
\end{figure}

Due to the open boundaries and the presence  of the edge fields $(h_{ L}, h_{  R})$, the spin profiles    $S^z_j= \langle \sigma_j^z \rangle /2$
in both the ground state and the midgap state differ from the bulk antiferromagnetic order close to the boundaries. For large enough $N$ we may write
\be
S^z_j= (-1)^j \frac{\sigma}{2}+ \Delta S^z(j),
\label{spinprofilesigma}
\ee
where
\be
\sigma=\pm \left(\Pi_{n= 1}^{\infty}  (\frac{1-q^{2n}}{1+q^{2n}})\right)^2, \; q=e^{-\gamma},
\ee
is the exact staggered magnetization of the XXZ chain in the thermodynamical limit and  $\Delta S^z(j)$ is the relative deviation with respect to the AF bulk profile.
Due to the gap in the bulk these deviations are expected to be   localized  close to both  the left and the right edges 
\be
 \Delta S^z(j) =  \Delta S^z_L(j) + \Delta S^z_R(j),
 \label{ALR}
 \ee
 where $ \Delta S^z_{L,R}(j) $ are localized close to $j=1$ and $j=N$
respectively  (i.e: $ \Delta S^z_{L,R}({N/2}) \sim e^{-N/2}$). This is indeed what we find. We plot in Fig.~\ref{DS_odd} our DMRG  results for $\Delta S^z_L(j)$ in both the ground state and the midgap state. 
We clearly observe an exponential localization of the relative spin accumulation for various values of $\Delta$ at constant boundary fields
$h_L=h_R=0.2$ (see insets of Fig.~\ref{DS_odd}). 

\section{Spin fractionalization}
The above spin accumulations, or depletion,  do not come as a surprise
and are expected due to the open boundaries and the presence of the edge fields. What is non trivial is that they correspond to a genuine spin fractionalization in both the ground state and the midgap state. As we shall now demonstrate, in the thermodynamical limit and for all $\Delta >1, h_{{ L},{ R}}$, there exist  fractionalized quarter spin operators associated with  each edge, ${\cal \hat S}^z_{ L}$ and ${\cal  \hat S}^z_{ R}$, which  have well defined fractional  eigenvalues
\be
 {\cal \hat S}^z_{{ L},{ R}}|g(e)\rangle = { \cal S}_{{ L},{R}}|g(e)\rangle, \; {\cal  S}^z_{ L,  R} =\pm \frac{1}{4}.
 \label{fracspindef}
 \ee
 In the basis $(|g\rangle, |e\rangle)$
the above fractional spin operators commute with each other, and anticommute with the spin flip operator, i.e: $[{\cal \hat S}^z_{ L}, { \hat S}^z_{ R}]=0$, $\{\tau, {\cal \hat S}^z_{ L,  R}\} =0$. Together, they  reconstruct the  z-component of the  total spin  $\hat S^z =\sum_{i=1}^N \hat S_i^z$, namely
 \be
\hat S^z ={\cal \hat S}^z_{ L} + {\cal { \hat  S}}^z_{R}.
\label{SzTfrac}
\ee 
Since the edge spin operators  have fractional spin $\pm 1/4$ one may verify that the $\hat S^z$  have eigenvalues  $0$ or $\pm 1/2$ depending on whether  $N$ is even or odd. For the fractional spin operators (\ref{fracspindef})  to describe sharp quantum observables in the subspace spanned by $(|g\rangle, |e\rangle)$, not only they have to average to  $\pm 1/4$ in both states,  but  also  their  variance must vanish in the thermodynamical limit, i.e:
 \bea
\langle { \hat S}^z_{{ L},{ R}}\rangle &=&  { \cal S}_{{ L},{ R}},
\label{fracspinaverage}
\eea
and 
\bea
  \delta {\cal S}^2_{{ L}, { R}}   &\jp{\equiv}&  \langle  ({\cal \hat S}^z_{{L}, { R}})^2 \rangle -({ \cal S}_{{ L},{ R}})^2=0,
\label{variance}
\eea
where the average $\langle ... \rangle$ is taken in each of the two states $(|g\rangle$ and $|e\rangle)$.

Following the authors of Refs.\cite{kivelsonschriefer,Jackiw} we define the fractional spin operators as their convolution with a decaying function $f(x)$, here we take $f(x)=e^{-\alpha x}$ to write
\bea
{\cal \hat S}^z_{ L} &=& \lim_{\alpha \rightarrow 0} \lim_{N \rightarrow \infty}  \sum_{j=1}^N f(j) \frac{\sigma^z_j}{2},\\
 {\cal \hat S}^z_{ R} &=& \lim_{\alpha \rightarrow 0} \lim_{N \rightarrow \infty}  \sum_{j=1}^N f(N+1- j)\frac{\sigma^z_j}{2},
 \label{fractional spins}
\eea
which   takes the limit $\alpha \rightarrow 0$ after the limit $N \rightarrow \infty$.
 We stress that the order of limits in (\ref{fractional spins}) is  important since by taking the limit $\alpha \rightarrow 0$  first,
 both ${ \cal S}^z_{{ L},{ R}}$ would identify with the total magnetization $S^z$. 

Due to the AF long range order it is convenient to distinguish between the contributions of the staggered part of the spin profile and that of the exponentially localized contributions
\be
{\cal  \hat S}^z_{{ L}} = -\frac{\sigma}{4} + {\Delta \cal  \hat S}^z_{{L}},\; {\cal  \hat  S}^z_{{ R}} = -\frac{\sigma}{4}  (-1)^N+ {\Delta \cal \hat  S}^z_{{ R}}.
 \label{FRrelation}
\ee
where the relative  accumulation operators are given by
\bea
 {\Delta \cal   \hat S}^z_{{ L}} &=& \lim_{\alpha \rightarrow 0} \lim_{N \rightarrow \infty}  \frac{1}{2}\sum_{j=1}^N    [ \sigma_j^z- \sigma (-1)^j] e^{-\alpha j}, \nonumber \\
  {\Delta \cal   \hat S}^z_{ { R}} &=& \lim_{\alpha \rightarrow 0} \lim_{N \rightarrow \infty} \frac{1}{2} \sum_{j=1}^N  [ \sigma_j^z- \sigma (-1)^j] e^{-\alpha (N+1-j)}.
  \label{fractional relative spins}
\eea
We have used the identity $\lim_{\alpha \rightarrow 0} \sum_{j=1}^{\infty}  (-1)^j e^{-\alpha j}= -\frac{1}{2}$.
Numerically (\ref{fractional relative spins}) are much easier to investigate since the relative spin accumulations (\ref{DS_odd}) are exponentially localized. In practice one may set $\alpha=0$  in (\ref{fractional relative spins})  provided the summation over  $j$ extends to the middle of the chain $j_{\rm max}=N/2$. Convergence is then expected to be of order $e^{-N/2}$. Before going further, it is worthy to point out that although
the relative accumulations  defined in Eq.(\ref{fractional relative spins}) have the same variance as the fractional spin operators
 (\ref{fractional spins}) they do not qualify as spin operators in the sense that  they do not anticommute with the spin flip operator 
$\tau$, i.e: $\{\Delta {\cal \hat S}_{{ L}, { R}} , \tau\} \neq 0$ \footnote{For instance   it would not couple to an external magnetic field penetrating smoothly near the left edge whereas ${\cal \hat S}_{{ L}}$ would.}. These relative accumulations
would have fractional eigenvalues which depend on the  anisotropy parameter $\Delta$.
Taking into account the AF long range order in the bulk is essential for the spin accumulations to have fractional eigenvalues $\pm 1/4$ independently of the model parameters as we shall see.

\begin{figure}
    \centering
    \includegraphics[width=0.5\textwidth]{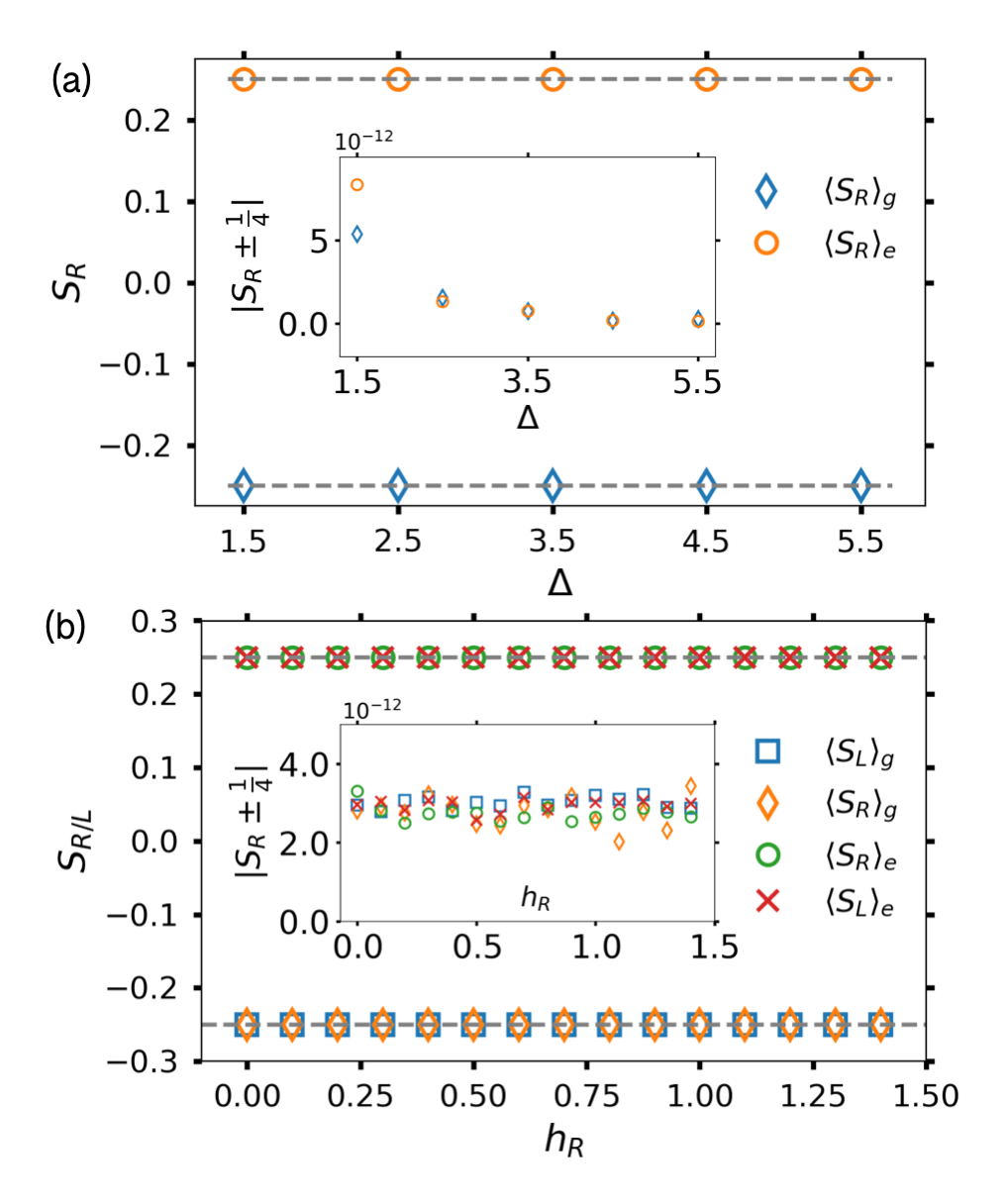}
    \caption{Edge spin accumulation $ {\cal S}_{L/R}= \langle {\cal \hat S}^z_{ L/R} \rangle $ in the ground state $|g\rangle$, with total spin $S^z=-\frac{1}{2}$, and in the midgap  state $|e\rangle$, with total spin $S^z=\frac{1}{2}$. The model parameters used are  (a) $N=1001$, $h_{R}=h_L=0.2$ with varying $\Delta$. (b) $N=1001$, $h_{ L}=0$ and $\Delta=3$ with varying $h_{ R}$. The dashed grey lines are on the curve $S=\pm 1/4$ to represent the expected value. Insets show the difference of numerical values from the expected value, which is on the order of the DMRG accuracy $\sim 10^{-12}$.}
    \label{Spin_accumulation_odd}
\end{figure}

\paragraph{ Spin $\pm \frac{1}{4}$ accumulations.}
We have computed, using extensive DMRG calculations,  the edge spin accumulations  ${\cal  S}_{ L}= \langle {\cal \hat S}^z_{ L} \rangle $ and  ${\cal  S}_{ R}= \langle {\cal \hat S}^z_{ R} \rangle $ in both the ground state and the midgap state  for a wide range of boundary fields $|h_{ L,  R}| < \Delta -1$ and  parameters $\Delta >1$. All together our results are consistent with an accumulation of  a spin ${\cal  S}_{ L,  R}= \pm 1/4$ at the two edges of the system
in both the ground state and the midgap state. Furthermore we verify explicitly that  these quarter spins reconstruct the total spin $S^z$, as given by Eq.(\ref{SzTfrac}), of the ground state and the midgap state 
for both  $N$ even and $N$ odd. We show here our results for an odd number of sites fixing $\Delta=3$ and an anisotropic edge fields configuration  $h_{ L}=0$ with varying $h_{ R}$ in the Figs.(\ref{Spin_accumulation_odd}).
To check that the quarter spins observed so far do not depend on the value of $\Delta > 1$, we also show
the spin accumulations fixing $h_L=h_R=0.2$ (in this case ${\cal  S}_{ L}={\cal  S}_{ R}$ thanks to the $\mathbb{P}$ symmetry)  and varying $\Delta$. More results are given in the Supplementary Materials.
  
\begin{figure}
    \centering
    \includegraphics[width=0.5\textwidth]{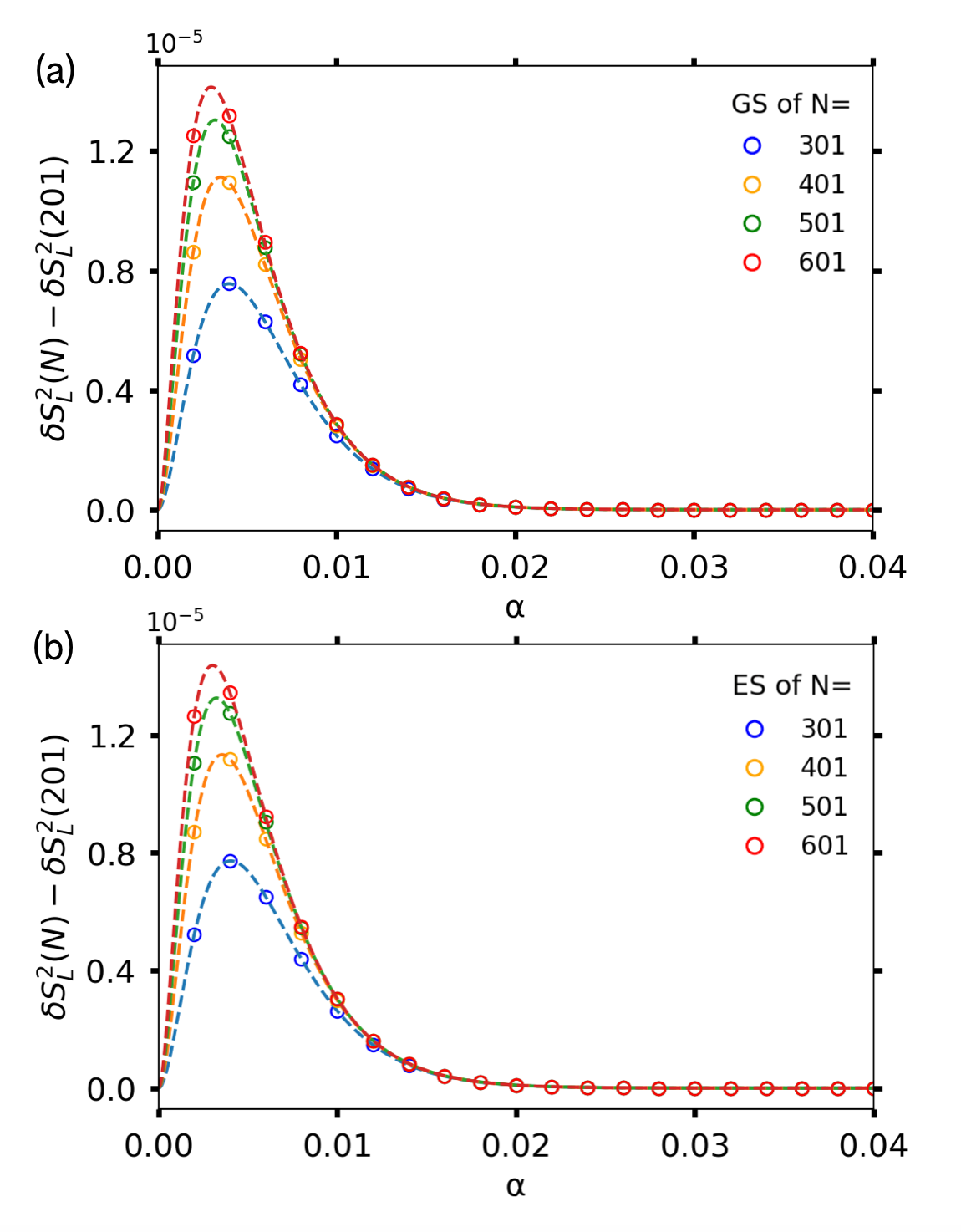}
    \caption{Edge spin variance $\delta S_{ L}^2(N,\alpha)$ in (a) the ground state (GS) $S^z=-\frac{1}{2}$ and (b) the midgap state (ES) $S^z=\frac{1}{2}$ with model parameters $\Delta=3$, $h_{ L}=h_{ R}=0.2$ and varying $N$.}
    \label{var}
\end{figure}

\paragraph{Variance.} We also calculated the spin variance to directly verify that the quarter spins found so far are sharp quantum observables.
To this end we define the spin variance at, say, the left edge for a finite system $N$ and cutoff $\alpha$ as 
\begin{equation}
    \delta {\cal S}^2_{L}(N,\alpha) = \langle {\cal S}_L^z(N,\alpha)^2\rangle -\langle S_{ L}^z(N,\alpha)\rangle^2,
\end{equation}
where the average is taken in either the ground state or the midgap state   $|g\rangle$ and  $|e\rangle$.  In the thermodynamic limit, the  variance 
as defined  in Eq.(\ref{variance}), is then obtained as 
\begin{equation}
    \delta S_{ L}^2 \equiv \lim_{\alpha\rightarrow 0}\lim_{N\to\infty}\delta S^2(N,\alpha).
\end{equation}
Taking the $N\rightarrow\infty$ is challenging and we circumvent this issue by assuming an ansatz relating $\delta S^2_{ L}(N,\alpha)$ and $\delta S^2_{ L}(\infty,\alpha)$ 
\begin{equation}
     \delta S^2_{ L}(N,\alpha)= \delta S^2(\infty,\alpha)-\frac{A}{\Delta}\alpha e^{-B\alpha N}.
\end{equation}
With the above ansatz  $\delta S^2_{ L} = \lim_{\alpha\rightarrow 0}S(\infty,\alpha)=0$, and we can hence  calculate $\delta S^2_{ L}$ without taking explicitly the thermodynamic limit.
We have verified  this ansatz by taking the difference of $\delta S^2_{ L}(N,\alpha)$ for different $N$'s. This is shown in Fig.\ref{var}, where one can see that the ansatz fits the data very well. The fitted parameter $B\approx 2$ is nearly independent of the boundary fields, while $A$ takes a non-universal value. 


In summary we find that in the low energy subspace spanned  by the ground state $|g\rangle$ and  the midgap state $|e\rangle$ one can assign to the left and the right edges a fractional spin state with eigenvalues $ {\cal S}_{{ L}, { R}}= \pm \frac{1}{4}$. On the basis of our results we find it safe to expect that this is to be the case irrespective of the anisotropy parameter $\Delta>1$ and the values of the edge fields $[h_{L,R}| < \Delta-1$.
Due to the zero variance of the 
fractional   spin operators (\ref{fractional spins}),  
 the quarter spins  ${\cal  S}_{{ L}, { R}}$ 
%
are  not  simple quantum averages of half-integers spins at different sites but rather   sharp quantum observables. 
The   orientations of these quarter spins depend on the boundary fields and on the parity of the number of sites $N$
in such a way that  (\ref{SzTfrac}) is satisfied in all ground states. 
Since the fractional spins at each edge are good quantum numbers we may then label the ground state and the midgap state  as 
$|g(e)\rangle =  |{\cal S}_{{ L}}, {\cal S}_{{ R}}\rangle$.
For odd $N$ spin chains these states
are given by $|\pm 1/4, \pm 1/4\rangle$ whereas for even chains
they are given by $|\pm 1/4, \mp 1/4\rangle$. One can easily verify  that the total spin is $S^z=\pm 1/2$ and $S^z=0$ for the odd and even cases.

We want to point out that the existence of a gap above the  ground state and the midgap state seems to be crucial for 
 the   quarter spins to be sharp quantum observables. Indeed, in the limit $\Delta\rightarrow 1$ where the mass gap goes to zero, we end up with  the $XXX$ Heisenberg chain. In this case  it was found in \cite{XXXmagpaper} that although the fractional spin $\pm 1/4$ exist in the ground state, their variance is not  zero and hence the fractional spins are not genuine quantum observables.

\section{Discussion}
The first natural question that arises is whether or not the quarter spins found so far survive in the higher excited states of the spectrum of the XXZ chain. 
However, excited states above the midgap state contain propagating spinons. In such case, even if a quarter spin can be defined {\it on average}, we do not expect its variance to be zero as found for the XXX spin chain with edge fields \cite{XXXmagpaper}. Another related question is whether these quarter spins survive edge fields higher than the critical value $h_c =\Delta-1$. In this regimes there are no midgap states \cite{skorik,kapustinxxz,xxzbound2019,Nassar_1998} but  we believe that   sharp  quarter spins exist in the ground state due to the existence of the spectral gap.

We shall end by commenting about the relation between the quarter spins found in this work with spontaneous symmetry breaking of the $\mathbb{Z}_2$ symmetry in the case of zero edge fields, i.e: $h_L=h_R=0$. In the limit of zero edge fields the two states $|g(e)\rangle$ become degenerate in the thermodynamical limit as the bound state energies 
(\ref{bsenergy}) vanish. Without loss of generality one may then choose $|g\rangle = |-1/4, -1/4\rangle$ for $N$ odd and $|g\rangle = |1/4, -1/4\rangle$ for $N$ even.
The two linear combinations $|\pm\rangle =(|g\rangle \pm |e\rangle)/\sqrt{2}$ are eigenstates of $\tau$, i.e: $\tau |\pm\rangle = \pm |\pm\rangle$. Since, in the same limit each of the two states $|g(e)\rangle$ are eigenstates of ${\hat S}^z_{L,R}$,  the fractional spin operators map the two states $|\pm\rangle $ onto each other, i.e: ${\hat S}^z_{L,R} |\pm\rangle = (-1)^N\frac{1}{4} |\mp\rangle$. 
Hence the fractional spin operators are zero energy modes (ZEM) in the basis $|g\rangle$ and $|e\rangle$. Notice that, since in this subspace ${\hat S}^z_{L,R}$ are not independent as  ${\hat S}^z_{L}= \pm {\hat S}^z_{R}$ in both states,  there exits only one ZEM say ${\hat S}^z_{L}$.
At this point it is worth mentioning that the  hamiltonian (\ref{XXZ}) displays the remarkable property, discovered by P. Fendley\cite{Fendley}, of having a strong zero energy mode $\Psi_F$  in the thermodynamical limit satisfying the following properties
 \bea
[\Psi_F, H]=0, \,\,\,  \{\Psi_F, \tau\}=0, \,\,\,
\Psi_F^2=1.
\label{FZEM}
 \eea
 The existence of the later operator insure that in the $N\rightarrow \infty$ limit the Hilbert space associated with the XXZ spin chain fractionalizes into two degenerated towers with eigenvalues $\tau=\pm 1$ which are mapped onto each other by the action of $\Psi_F$. We may therefore conclude that 
 when {\it projected} in the low energy subspace spanned by the ground state and the midgap state the Fendley operator identifies with the fractional spin operator
 \be
\Psi_F \equiv 4 {\hat S}^z_{L}.
\ee
 Of course, since we do not expect
 a quarter fractional spin to be sharp in all the excited states, 
 ${\hat S}^z_{L}$ is not a strong ZEM
 in contrast with the Fendley operator $\Psi_F$ but rather a soft
 ZEM. We finally notice that, following the same  lines of arguments as given above, the fractional spin $ {\hat S}^z_{L}$ is also a  soft ZEM on the two lines $h_L=h_R$ for $N$ even and $h_L=-h_R$ for $N$ odd where the symmetries $\mathbb{P} \circ \mathbb{Z}_2$ and $\mathbb{P}$ are spontaneously broken. It would interesting to know if a strong zero mode similar to (\ref{FZEM}) exists on these two symmetric lines. We hope that the quarter spins found in this work could be probed in experiments using ultra cold atoms in optical lattices \cite{Demlerultracold}.

\acknowledgements{This work is is partially supported by the Air Force Office of Scientific Research under Grant No.~FA9550-20-1-0136 (J.L.,J.H.P.), NSF Career Grant No.~DMR-1941569 (J.H.P.), and the Alfred P.~Sloan Foundation through a Sloan Research Fellowship (J.H.P.).}
 
\bibliography{refXXZ}
\appendix
\begin{widetext}
\section{Even N DMRG}

\begin{figure}
    \centering
    \includegraphics[width=0.45\textwidth]{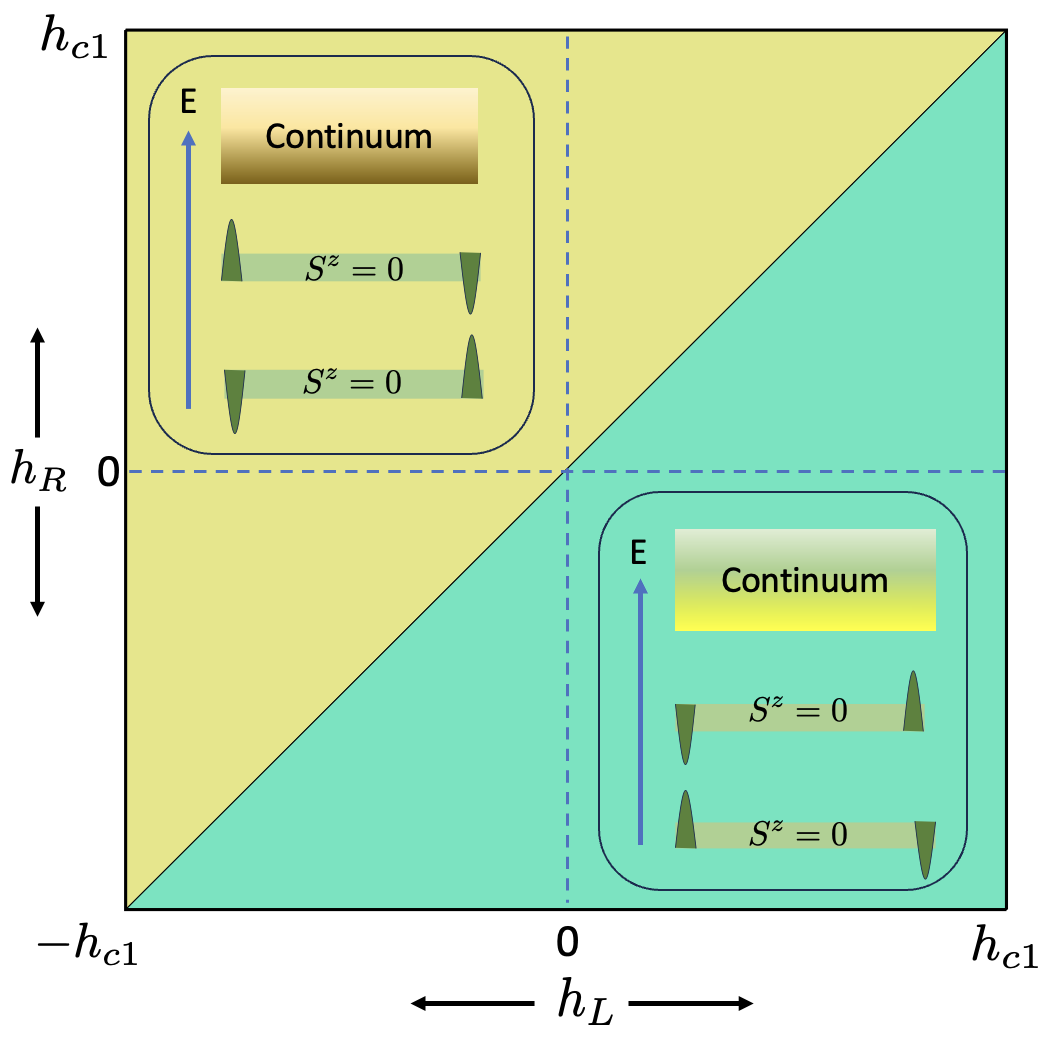}
    \caption{Ground state phase diagram of the XXZ model with edge fields smaller than the critical field $|h_{L,R}| < h_c=\Delta-1$  and for an even number of sites. The ground state as well as the first  excited state which is a midgap state have total spin $S^z=0$. Both states host fractional quarter spins ${\cal S}_{L,R}=\pm 1/4$ at both edges of the chain. These fractional spins  are sharp quantum observables which reconstruct the total spin of each state $S^z={\cal S}_{L}+{\cal S}_{R} =0$. The $\pm 1/4$ quarter spins are depicted  by triangles pointing  upward and downward respectively. On the separatrix $h_L-h_R=0$ the two states are degenerate and the edge spin operator become a zero energy mode. }
    \label{pdeven}
\end{figure}

\subsection{Edge spin accumulation}
The lowest-energy state of the XXZ Hamiltonian defined in Eqn.(\ref{XXZ}) can be solved numerically in a matrix product (MPS) form by the density matrix renormalization group (DMRG) method.
The magnetization at site $j$ can be computed as $S^z_j = \frac{1}{2}\langle \psi| \sigma^z_x |\psi\rangle$. Also, we define an Ansatz for the spin profile as 
\begin{equation}
\begin{split}
    S^z_j =&\Delta S^z(j) +\frac{1}{2}\sigma(\Delta) (-1)^x
\end{split}
\end{equation}
with $\sigma(\Delta)$ is the bulk staggered magnetization for a periodic $XXZ$ chain with anisotropy $\Delta$. The numerical results for the validity of fitting in the lowest two excited states are shown in Figs.(\ref{DS_even},\ref{DSL_even}) for even cites and in Figs.(\ref{DS_odd}) for the odd sites.

With the definition of the edge spin accumulation as 
\begin{equation}
    \label{A_S}
    {\cal S}_{L} = \lim_{\alpha \to 0}\lim_{N\to\infty}\sum_j e^{-\alpha x}S^z(j,N)=\lim_{\alpha \to 0}\lim_{N\to\infty}S^z_L(N,\alpha),
\end{equation}
then ${\cal S}_{L/R}=\pm\frac{1}{4}$ in both of the two lowest energy states, and they sum up to ${\cal S}_L+{\cal S}_R = S^z$. Noticing, that the non-converging part is the bulk magnetization if we want to define the edge spin accumulation on a finite system, we introduce the quantity $S_{\alpha} = \pm\frac{1}{4} = \sum_x \Delta S_\alpha^z(x)\pm \frac{1}{2}\sigma(\Delta)$ with $\alpha = L,R$. The edge spin accumulation is shown in Fig.(\ref{Spin_accumulation_even}) for a chain with even number of sites and Fig.(\ref{Spin_accumulation_odd}) for a chain with odd number of sites.

\begin{figure}[H]
    \centering
    \includegraphics[width=\textwidth]{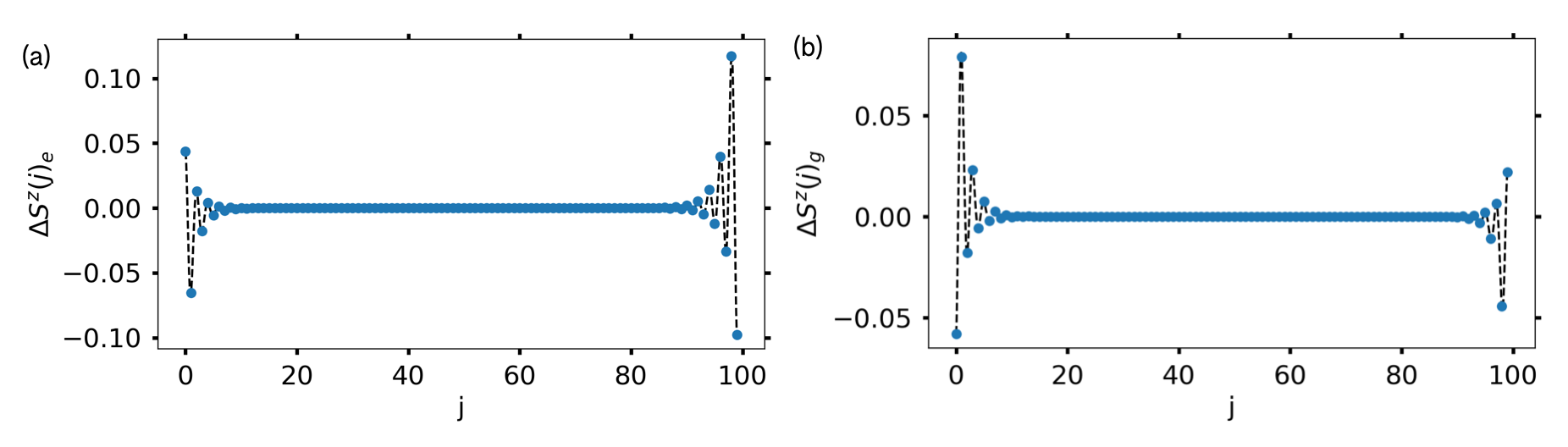}
    \caption{Spin profile $\Delta S^z(j)$ and the fitting of ansatz (a) in the ground  state $|e\rangle$  (b) in the midgap  state $|g\rangle$ with model parameters $N=100$, $h_L=0.1,h_R=0.5$ and for $\Delta=3$.}
    \label{DS_even}
\end{figure}
\begin{figure}[H]
    \centering
    \includegraphics[width=\textwidth]{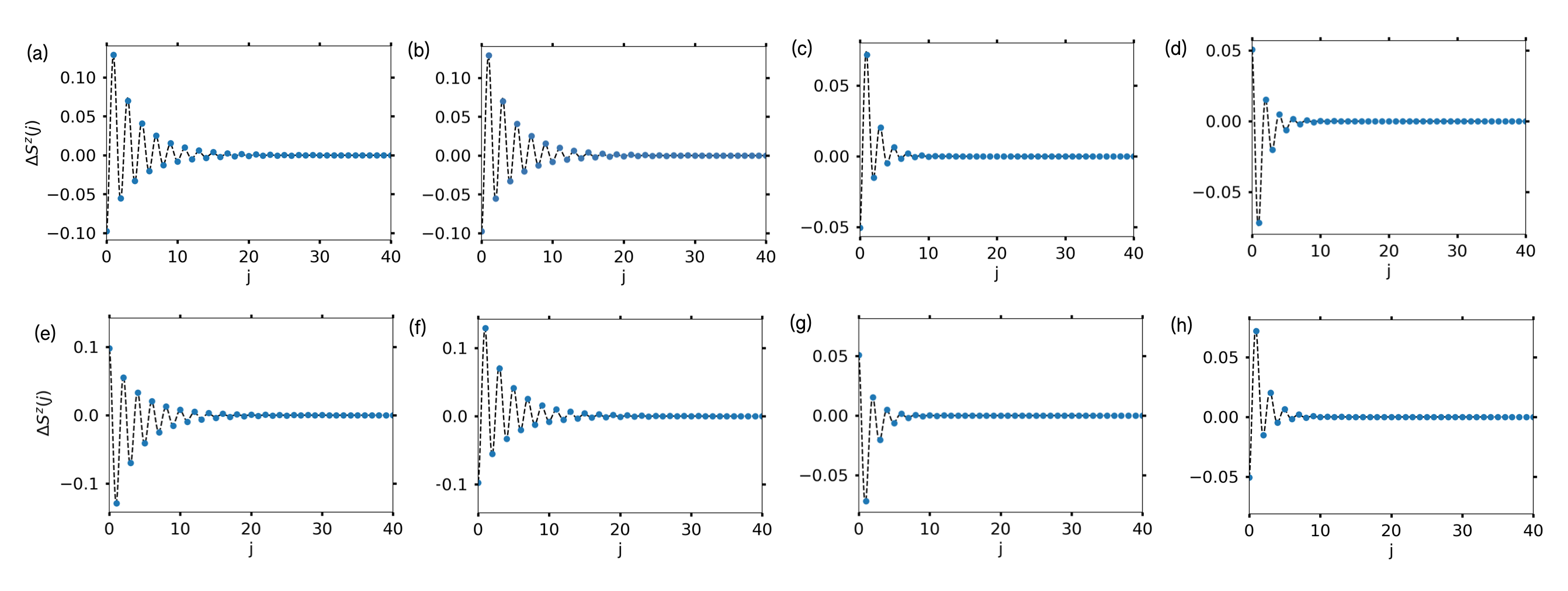}
    \caption{Spin profile on the left hand side $\Delta S^z_L(j)$ of the two degenerated ground states without the presence of magnetic field $h_L=h_R=0$ for system size $N=1001$ and (a,b) $\Delta=2$ and (c,d) $\Delta=3$. Spin profile on the left hand side $\Delta S^z_L(j)$ of the two degenerated ground states without the presence of magnetic field $h_L=h_R=0$ for system size $N=1000$ and (e,f) $\Delta=2$ and (g,h) $\Delta=3$.}
    \label{DSL_even}
\end{figure}

\subsection{Edge Spin Variance}

Defining the spin variance operator at the edge for a finite system $N$ and cutoff $\alpha$ as 
\begin{equation}
    \delta S^2_L(N,\alpha) = \langle S_L^z(N,\alpha)^2\rangle -\langle S_L^z(N,\alpha)\rangle^2.
\end{equation}
The thermodynamic spin variance is defined through the same limit as in Eq.(\ref{A_S}), and the condition that the fractional spin ${\cal S}_{L/R}$ is a sharp quantum observable is that the variance vanishes:

\begin{equation}
    \delta S^2_L \equiv \lim_{\alpha\to\infty}\lim_{N\to\infty}\delta S_L^2(N,\alpha)=0.
\end{equation}

Taking the limit $N\rightarrow\infty$ is challenging, and we circumvent this issue by assuming an ansatz relating $\delta S^2_L(N,\alpha)$ and $\delta S^2_L(\infty,\alpha)$ 

\begin{equation}
     \delta S^2_L(N,\alpha)= \delta S_L^2(\infty,\alpha)-\frac{A}{\Delta}\alpha e^{-B\alpha L}.
\end{equation}

Then, $\delta S^2_L = \lim_{\alpha\rightarrow0}S(\infty,\alpha)$, and we can calculation the value of $\delta S^2_L$  in the thermodynamic limit. We verify this ansatz by taking the difference of $\delta S^2_L(N,\alpha)$ for different $N$'s.
This is shown in Fig.(\ref{varS_even}) for even chain and Fig.(\ref{var}) for odd chain, where one can see that the ansatz fits the data very well.

Therefore, the thermodynamic limit of the variance does vanish, $ \delta S^2=0$, and the edge spin is indeed a well-defined quantum observable.

\begin{figure}
    \centering
    \includegraphics[width=0.6\textwidth]{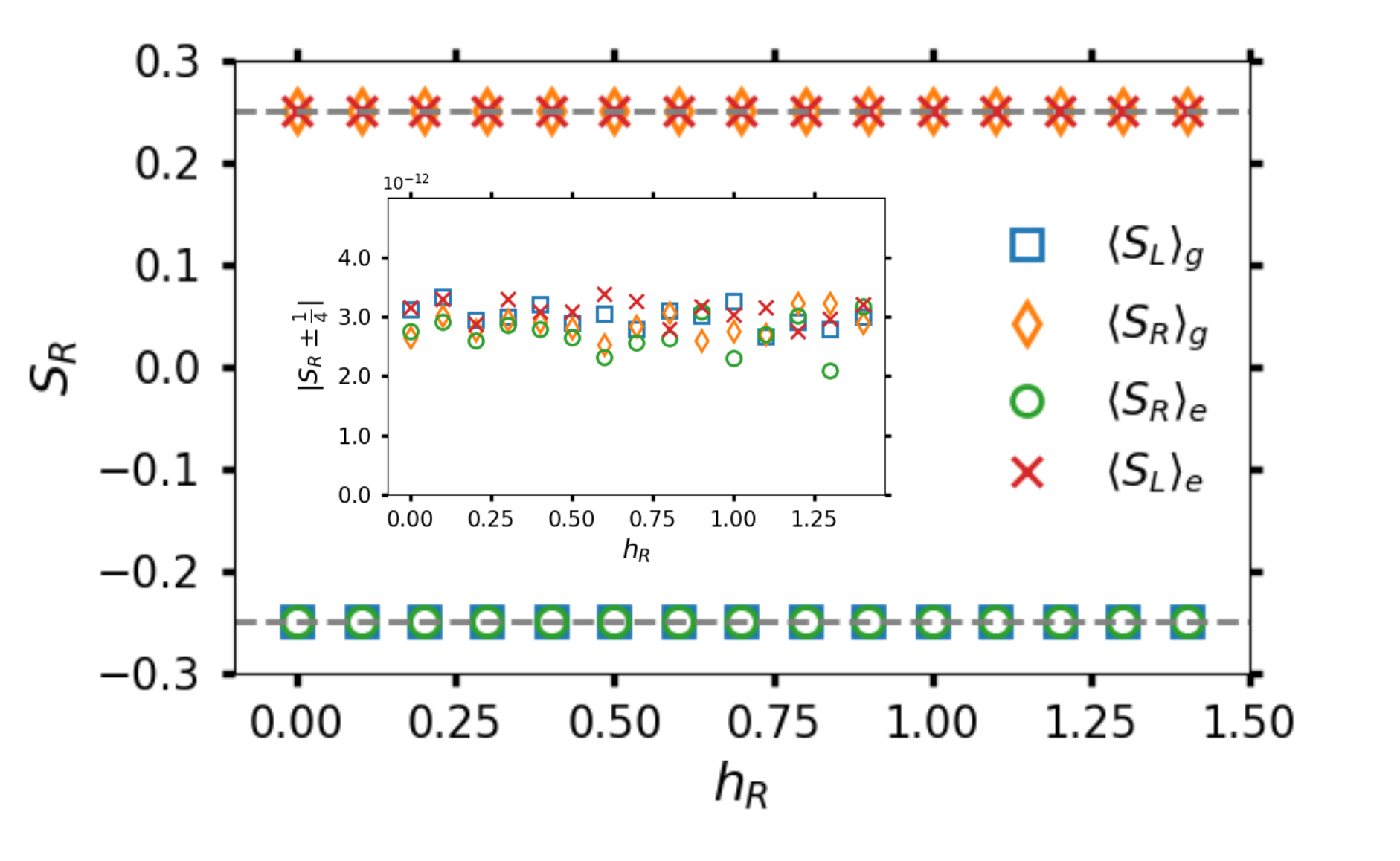}
    \caption{Edge spin accumulation ${\cal \hat S}_{\cal L/\cal R}= \langle {\cal \hat S}^z_{\cal L/\cal R} \rangle $ in the ground state $|g\rangle$, and in the midgap  state $|e\rangle$. The model parameters used are  $N=1000$, $h_{\cal L}=0$ and $\Delta=3$ with varying $h_{\cal R}$.  }
    \label{Spin_accumulation_even}
\end{figure}
\begin{figure}
    \centering
    \includegraphics[width=0.8\textwidth]{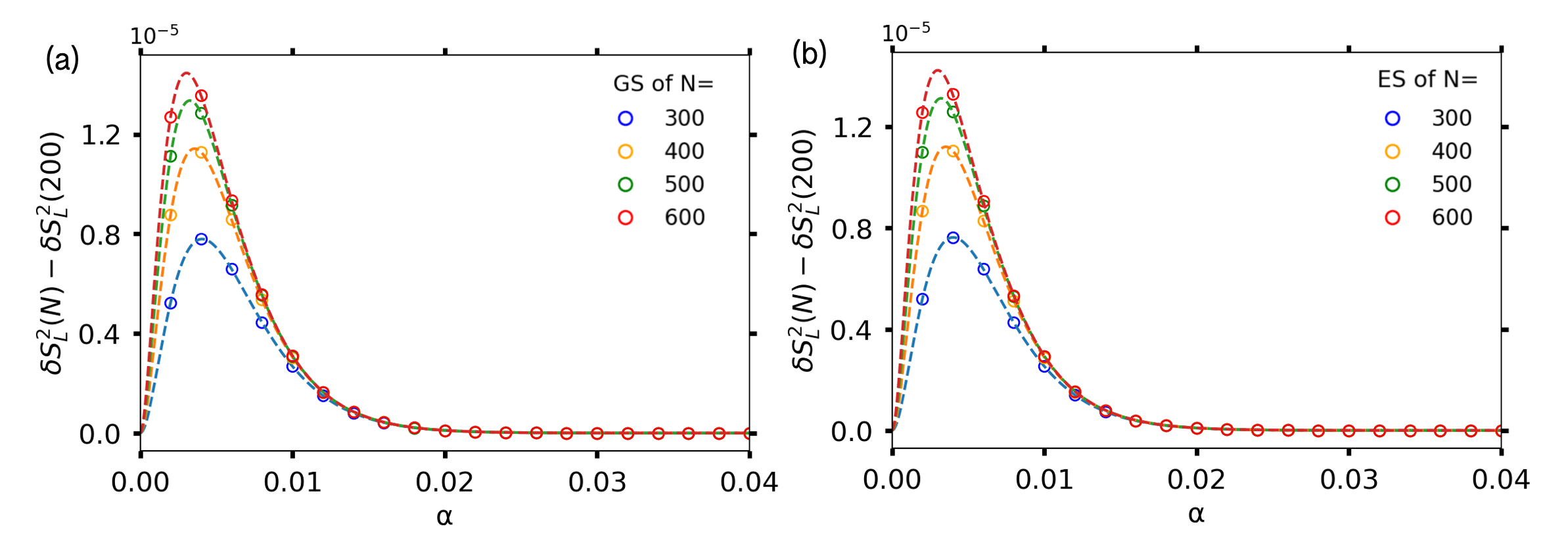}
    \caption{Edge spin variance $\delta S_{\cal L}^2(N,\alpha)$ in (a) the ground state $S^z=-\frac{1}{2}$ and (b) the midgap state $S^z=\frac{1}{2}$ with model parameters $\Delta=3$, $h_{\cal L}=h_{\cal R}=0.2$ and varying $N$.}
    \label{varS_even}
\end{figure}

\section{Bethe Ansatz}
\subsection{Hamiltonian}
Recalling the Hamiltonian of the system:

\bea
\label{hamiltonian}
H=\sum_{j=1}^{N-1} \left[\sigma^x_j\sigma^x_{j+1}+\sigma^y_j\sigma^y_{j+1}+\Delta (\sigma^z_j\sigma^z_{j+1}-1)\right]\nonumber\\+h_L\sigma^z_1+h_R\sigma^z_N,
\eea

where $h_L,h_R$ are magnetic fields at the left and the right edges respectively. We can introduce new parameters $\gamma$, $h_{c1}$, $h_{c2}$ such that

\bea \Delta=\cosh \gamma, \gamma>0,\;\;\;\;\; h_{c1}=\Delta-1, \;\;\;\;\; h_{c2}=\Delta+1 \eea

The Bethe equations can be obtained by following the method of coordinate or algebraic Bethe ansatz \cite{Alcaraz,Sklyanin,ODBA}. One obtains the following Bethe equations for the reference state with all spin up

\bea\label{be1}
\left(\frac{\sin\frac{1}{2}(\lambda_j-i\gamma)}{\sin\frac{1}{2}(\lambda_j+i\gamma)}\right)^{2N}\prod_{\alpha}^{L,R}\left(\frac{\sin\frac{1}{2}(\lambda_j+i\gamma(1+\epsilon_{\alpha}))}{\sin\frac{1}{2}(\lambda_j+i\gamma(1+\epsilon_{\alpha}))}\right)\nonumber\\=\prod_{\sigma=\pm}\prod_{k=1}^M\left(\frac{\sin\frac{1}{2}(\lambda_j+\sigma\lambda_k-2i\gamma)}{\sin\frac{1}{2}(\lambda_j+\sigma\lambda_k+2i\gamma)} \right)
\eea

where \bea h_\alpha= -\sinh\gamma\coth(\frac{\epsilon_{\alpha}\gamma}{2}), \;\;\;   \epsilon_\alpha=\tilde{\epsilon}_{\alpha}+i\delta_\alpha \frac{\pi}{\gamma},\nonumber\\ \delta_{\alpha} =\begin{cases}  1 & |h_{\alpha}|<\sinh\gamma \\ 0 & |h_{\alpha}|>\sinh\gamma \end{cases} \eea

 Note that $h_{c1}<\sinh\gamma<h_{c2}$. The Bethe equations for reference state with all spin down can be obtained by the transformation $\epsilon_{\alpha}\rightarrow -\epsilon_{\alpha}$ \cite{Sklyanin}. The energy of a state described by the set of Bethe roots $\lambda_j$ is given by

\bea \label{energy}E=\frac{1}{2}\left[(N-1)\cosh\gamma+h_L+h_R\right]\nonumber\\-2\sinh\gamma\sum_{j=1}^M\frac{\sinh\gamma}{\cosh\gamma-\cos\lambda_j}
\eea

The boundary magnetic fields break the $\mathbb{Z}_2$ spin flip symmetry. Under the spin flip of all the sites, the bulk remains invariant but the boundary terms remain invariant only after the direction of both the magnetic fields is reversed, hence we have the following isometry 
\bea \label{z2} \prod_{i=1}^{N}\sigma^{x}_i H \sigma^{x}_i, \;\;h_L\rightarrow-h_L, \;\; h_R\rightarrow-h_R.\eea

\subsection{Bethe Solution}

In this section we construct the ground states and the boundary excitations with the lowest energy in each of the four sub-phases $A_{j=(1,2,3,4)}$, corresponding to the domains of the boundary fields $(0\le h_{L}\le h_{c1},0\le h_{R}\le h_{c1}), (0\ge h_{L} \ge -h_{c1},0\le h_{R}\le h_{c1}) ,(0\ge h_{L} \ge -h_{c1},0\ge h_{R}\ge -h_{c1})$ and $(0\le h_{L} \le h_{c1},0\ge h_{R}\ge -h_{c1})$ respectively.

\subsubsection{Region $A_1$: odd number of sites}
 The region $A_1$ corresponds to the following values of the boundary magnetic fields: $0<h_L,h_R< h_{c1}$. This corresponds to $\epsilon_{\alpha}=-\tilde{\epsilon}_{\alpha}+i\pi$, with $\tilde{\epsilon}_{\alpha}<1$, $\alpha=L,R$.

\smallskip

First consider the state with all real $\lambda$, which take values between $(-\pi,\pi]$. Applying logarithm to \ref{be1} we obtain 

\bea\label{logbea11}
2N \varphi(\lambda_j,1)-\sum_{\alpha=L,R}\varphi(\lambda_j,1-\tilde{\epsilon}_{\alpha})+\varphi(\lambda_j,1)+\varphi^{\prime}(\lambda_j,1)=2\pi I_j+ \sum_{\sigma=\pm}\sum_{k\neq j}\varphi(\lambda_j+\sigma \lambda_k,2).\nonumber\\
\eea

where \bea \varphi(x,y)=\ln \left(\frac{\sin \frac{1}{2}(x-i\gamma y)}{\sin \frac{1}{2}(x+i\gamma y)}\right), \;\;\;\;  \varphi^{\prime}(x,y)=\ln \left(\frac{\cos \frac{1}{2}(x-i\gamma y)}{\cos \frac{1}{2}(x+i\gamma y)}\right).\eea

We define the counting function $\nu(\lambda)$ such that $\nu(\lambda_j)=I_j$. Differentiating \ref{logbea11} and using $\frac{d}{d\lambda}\nu(\lambda)=\rho(\lambda)$, we obtain

\bea\label{logbea12}
(2N+1) a(\lambda,1)-\sum_{\alpha=L,R}a(\lambda-\pi,1-\tilde{\epsilon}_{\alpha})+a(\lambda-\pi,1)-2\pi\delta(\lambda)-2\pi\delta(\lambda-\pi)=2\pi\rho(\lambda)+\sum_{\sigma=\pm}\int a(\lambda+\sigma \mu,2)\rho(\mu)d\mu,\nonumber\\ \eea

where we have removed the solutions $\lambda=0,\pi$ as they lead to a vanishing wavefunction \cite{skorik}. Here

\bea a(x,y)=\frac{\sinh(\gamma y)}{\cosh(\gamma y)-\cos(\lambda)}.\eea

The above equation can be solved by applying Fourier transform

\bea f(x)= \sum_{k=-\infty}^{\infty}\hat{f}(\omega)e^{i\omega x}, \;\;\;\;\;\;\;\; \hat{f}(\omega)=\frac{1}{2\pi}\int_{-\pi}^{\pi}f(x)e^{-i\omega x}dx. \eea

Using $\hat{a}(\omega,y)=e^{-\gamma y |\omega|}$, we obtain the following density distribution for the state with all real roots

\bea \label{denodda11}\hat{\rho}_{\ket{\frac{1}{2}}_{A_1}}(\omega)=\frac{(2N+1)e^{-\gamma|\omega|} + (-1)^{\omega}e^{-\gamma|\omega|}-(1+(-1)^{\omega})-\sum_{\alpha=L,R}(-1)^{\omega}e^{-\gamma (1-\tilde{\epsilon}_{\alpha})|\omega|}}{4\pi(1+e^{-2\gamma|\omega|})}. \eea

The reason for the subscripts will become evident when we find the spin $S^z$ of the state. The number of Bethe roots can be obtained by using the relation 
\bea \label{nroots1}M= \int_{-\pi}^{\pi} \rho(\lambda)d\lambda. \eea 

The total spin $S^z$ of the state can be found using the relation $S^z=\frac{N}{2}-M$. Using \ref{denodda11} in the above relations we find that the total spin $S^z$ of the state described by the distribution $\hat{\rho}_{\ket{\frac{1}{2}}_{A_1}}(\omega)$ is $S^z =\frac{1}{2}$.  We denote this state by $\ket{\frac{1}{2}}_{A_1}$.

\smallskip

By starting with the Bethe equations corresponding to all spin down reference state we have

\bea\label{logbea13}
(2N+1) a(\lambda,1)-\sum_{\alpha=L,R}a(\lambda-\pi,1+\tilde{\epsilon}_{\alpha})+a(\lambda-\pi,1)-2\pi\delta(\lambda)-2\pi\delta(\lambda-\pi)=2\pi\rho(\lambda)+\sum_{\sigma=\pm}\int a(\lambda+\sigma \mu,2)\rho(\mu)d\mu\nonumber\\ \eea

Following the same procedure as above, we obtain the following distribution for a state with all real $\lambda$ 

\bea \label{denodda12}\hat{\rho}_{\ket{-\frac{1}{2}}_{A_1}}(\omega)=\frac{(2N+1)e^{-\gamma|\omega|} + (-1)^{\omega}e^{-\gamma|\omega|}-(1+(-1)^{\omega})-\sum_{\alpha=L,R}(-1)^{\omega}e^{-\gamma (1+\tilde{\epsilon}_{\alpha})|\omega|}}{4\pi(1+e^{-2\gamma|\omega|})}. \eea

The total spin $S^z$ of this state is $S^z=-\frac{1}{2}$. We denote this state by $\ket{-\frac{1}{2}}_{A_1}$.

\smallskip

Using \ref{energy} we can calculate the energy difference between the two states $\ket{\frac{1}{2}}_{A_1}$ and $\ket{-\frac{1}{2}}_{A_1}$.

We have \bea \label{diffena11}E_{\ket{\frac{1}{2}}_{A_1}}-E_{\ket{-\frac{1}{2}}_{A_1}}= h_L+h_R-2\sinh\gamma\sum_{\alpha=L,R}\int_{-\pi}^{\pi}a(\lambda,1)\;\delta\rho_{\tiny{\ket{\frac{1}{2}},\ket{-\frac{1}{2}}}}(\lambda)d\lambda
,\eea

where $\delta\rho_{\tiny{\ket{\frac{1}{2}},\ket{-\frac{1}{2}}}}(\lambda)$ is the difference in the density distributions of the states $\ket{\frac{1}{2}}_{A_1}$ and $\ket{-\frac{1}{2}}_{A_1}$. The expression \ref{diffena11} can be written as
 
 \bea  E_{\ket{\frac{1}{2}}_{A_1}}-E_{\ket{-\frac{1}{2}}_{A_1}}= h_L+h_R+4\pi\sinh\gamma\sum_{\omega=-\infty}^{\infty} \hat{a}(\omega,1)\Delta\hat{\rho}_{\tiny{\ket{\frac{1}{2}},\ket{-\frac{1}{2}}}}(\omega). \eea

 Using \ref{denodda11} and \ref{denodda12} in the above expression we obtain
 \bea  E_{\ket{\frac{1}{2}}_{A_1}}-E_{\ket{-\frac{1}{2}}_{A_1}}= h_L+h_R+ \sinh\gamma\sum_{\alpha=L,R} \sum_{\omega=-\infty}^{\infty}(-1)^{\omega}\; \frac{\sinh(\gamma \tilde{\epsilon}_{\alpha}|\omega|)}{\cosh(\gamma\omega)}e^{-\gamma|\omega|}, \eea
 
which can be written as \bea E_{\ket{\frac{1}{2}}_{A_1}}-E_{\ket{-\frac{1}{2}}_{A_1}}=m_L+m_R,\eea where

\bea m_{\alpha}=h_{\alpha}+ \sinh\gamma \sum_{\omega=-\infty}^{\infty}(-1)^{\omega}\; \frac{\sinh(\gamma \tilde{\epsilon}_{\alpha}|\omega|)}{\cosh(\gamma\omega)}e^{-\gamma|\omega|}. \eea

The ground state is $\ket{-\frac{1}{2}}_{A_1}.$

 \subsubsection{Region $A_1$: Even number of sites}
 
The Bethe equations corresponding to all spin up reference state have two boundary string solutions $\lambda_{bs \alpha}=\pi+\pm i\gamma(1-\tilde{\epsilon}_{\alpha})$, $\alpha=L,R$ . Adding either of these two boundary strings to the Bethe equations \ref{be1} and taking logarithm we obtain

\bea \nonumber 2N \varphi(\lambda_j,1)-\sum_{\alpha=L,R}\varphi(\lambda_j-\pi,1-\tilde{\epsilon}_{\alpha})+\varphi(\lambda_j,1)+\varphi^{\prime}(\lambda_j,1)-\varphi(\lambda,(3-\tilde{\epsilon}_{\beta}))-\varphi(\lambda,(1+\tilde{\epsilon}_{\beta}))\nonumber\\=2\pi I_j+ \sum_{\sigma=\pm}\sum_{k\neq j}\varphi(\lambda_j+\sigma \lambda_k,2),
\eea

where $\beta$ is either $L$ or $R$. Differentiating the above equation with respect to $\lambda$ and taking the Fourier transform we obtain

\bea\label{denbounda11} \tilde{\rho}_{\ket{0}_{\beta A_1}} (\omega)=\tilde{\rho}_{\ket{\frac{1}{2}}_{A_1}}(\omega)+\Delta\tilde{\rho}_{\beta} (\omega),\eea

where

\bea \Delta\tilde{\rho}_{\beta} (\omega)=-\frac{1}{4\pi}(-1)^{\omega}\frac{e^{-\gamma(3-\tilde{\epsilon}_{\beta})|\omega|}+e^{-\gamma(1+\tilde{\epsilon}_{\beta})|\omega|}}{1+e^{-2\gamma|\omega|}}. \label{denbs}
\eea

The spin of the state containing this boundary string can be calculated using $S^z=\frac{N}{2}-M$, where
\bea \label{nroots2}M=1+\int_{-\pi}^{\pi}\rho_{\ket{0}_{\beta A_1}}(\lambda) d\lambda. \eea

We obtain $S^z_{\ket{0}_{\beta A_1}}=0$, $\beta=L,R$.  Hence there are two states with $S^z=0$ that correspond to the presence of the boundary strings $\lambda_{bsL}$ and $\lambda_{bsR}$.  

\smallskip

The energy of the boundary string can be calculated using \ref{energy}.  We have 

\bea \label{boundena1}
E_{\lambda_{bs\beta}}= -\frac{2\sinh^2\gamma}{\cosh\gamma+\cosh\gamma(1-\tilde{\epsilon}_{\beta})}-2\sinh\gamma\int_{-\pi}^{\pi}a(\lambda-\pi,1)\Delta\rho_{\beta}(\lambda)d\lambda.
\eea

Using \ref{denbounda11} and evaluating the integral one obtains,

\bea E_{\lambda_{bs\beta}}=-\frac{2\sinh^2\gamma}{\cosh\gamma+\cosh\gamma(1-\tilde{\epsilon}_{\beta})}+\sinh\gamma\sum_{\omega}(-1)^{\omega}e^{-2\gamma|\omega|}\frac{\cosh(\gamma(1-\tilde{\epsilon}_{\beta})\omega)}{\cosh(\gamma\omega)}=-m_{\beta} \label{bsenapp}.\eea

Hence the ground state is either $\ket{0}_{L, A_1}$ or $\ket{0}_{R, A_1}$ depending on the values of $h_L,h_R$.

\subsubsection{$A_2$: Odd number of sites}

The region $A_2$ corresponds to the following values of the boundary magnetic fields: $0<h_R< h_{c1}$, $-h_{c1}<h_L<0$. In this region the logarithmic form of the Bethe equations can be obtained from \ref{logbea11} by the transformation $\tilde{\epsilon}_L\rightarrow -\tilde{\epsilon}_L$. We have 

\bea
(2N+1) a(\lambda,1)-a(\lambda-\pi,1-\tilde{\epsilon}_{R})-a(\lambda-\pi,1+\tilde{\epsilon}_{L})+a(\lambda-\pi,1)-2\pi\delta(\lambda)-2\pi\delta(\lambda-\pi)\nonumber \\=2\pi\rho(\lambda)+\sum_{\sigma=\pm}\int a(\lambda+\sigma \mu,2)\rho(\mu)d\mu. \label{logbeA2} \eea

Taking Fourier transform we obtain

\bea \label{denodda21}\hat{\rho}_{\ket{\frac{1}{2}}_{A_2}}(\omega)=\frac{(2N+1)e^{-\gamma|\omega|} + (-1)^{\omega}e^{-\gamma|\omega|}-(1+(-1)^{\omega})-(-1)^{\omega}e^{-\gamma (1-\tilde{\epsilon}_{R})|\omega|}-(-1)^{\omega}e^{-\gamma (1+\tilde{\epsilon}_{L})|\omega|}}{4\pi(1+e^{-2\gamma|\omega|})}. \eea

 The number of Bethe roots can be obtained by using the relation \bea \label{nroots12}M= \int_{-\pi}^{\pi} \rho(\lambda)d\lambda. \eea 

The total spin $S^z$ of the state can be found using the relation $S^z=\frac{N}{2}-M$. Using \ref{denodda21} in the above relations we find that the total spin $S^z$ of the state described by the distribution $\hat{\rho}_{\ket{\frac{1}{2}}_{A_2}}(\omega)$ is $S^z =\frac{1}{2}$.  We denote this state by $\ket{\frac{1}{2}}_{A_2}$.

By starting with the Bethe equations corresponding to all spin down reference state we have

\bea
(2N+1) a(\lambda,1)-a(\lambda-\pi,1+\tilde{\epsilon}_{R})-a(\lambda-\pi,1-\tilde{\epsilon}_{L})+a(\lambda-\pi,1)-2\pi\delta(\lambda)-2\pi\delta(\lambda-\pi)\nonumber\\=2\pi\rho(\lambda)+\sum_{\sigma=\pm}\int a(\lambda+\sigma \mu,2)\rho(\mu)d\mu.  \label{logbea22} \eea

Following the same procedure as above, we obtain the following distribution for a state with all real $\lambda$ 

\bea \label{denodda22}\hat{\rho}_{\ket{-\frac{1}{2}}_{A_1}}(\omega)=\frac{(2N+1)e^{-\gamma|\omega|} + (-1)^{\omega}e^{-\gamma|\omega|}-(1+(-1)^{\omega})-(-1)^{\omega}e^{-\gamma (1+\tilde{\epsilon}_{R})|\omega|}-(-1)^{\omega}e^{-\gamma (1-\tilde{\epsilon}_{L})|\omega|}}{4\pi(1+e^{-2\gamma|\omega|})}. \eea

The total spin $S^z$ of this state is $S^z=-\frac{1}{2}$. We denote this state by $\ket{-\frac{1}{2}}_{A_2}$.

Using \ref{energy} we can calculate the energy difference between the two states $\ket{\frac{1}{2}}_{A_2}$ and $\ket{-\frac{1}{2}}_{A_2}$.

We have \bea \label{diffena21}E_{\ket{\frac{1}{2}}_{A_2}}-E_{\ket{-\frac{1}{2}}_{A_2}}= -h_L+h_R-2\sinh\gamma\sum_{\alpha=L,R}\int_{-\pi}^{\pi}a(\lambda,1)\;\delta\rho_{\tiny{\ket{\frac{1}{2}},\ket{-\frac{1}{2}}}}(\lambda)d\lambda,
\eea

where $\delta\rho_{\tiny{\ket{\frac{1}{2}},\ket{-\frac{1}{2}}}}(\lambda)$ is the difference in the density distributions of the states $\ket{\frac{1}{2}}$ and $\ket{-\frac{1}{2}}$. The expression \ref{diffena21} can be written as
 
 \bea  E_{\ket{\frac{1}{2}}_{A_2}}-E_{\ket{-\frac{1}{2}}_{A_2}}= -h_L+h_R+4\pi\sinh\gamma\sum_{\omega=-\infty}^{\infty} \hat{a}(\omega,1)\Delta\hat{\rho}_{\tiny{\ket{\frac{1}{2}},\ket{-\frac{1}{2}}}}(\omega). \eea

 Using \ref{denodda21} and \ref{denodda22} in the above expression we obtain
 \bea  E_{\ket{\frac{1}{2}}_{A_2}}-E_{\ket{-\frac{1}{2}}_{A_2}}= -h_L+h_R+ \sinh\gamma (-1)^{\omega}\; \frac{\sinh(\gamma \tilde{\epsilon}_{R}|\omega|)}{\cosh(\gamma\omega)}e^{-\gamma|\omega|}-\sinh\gamma (-1)^{\omega}\; \frac{\sinh(\gamma \tilde{\epsilon}_{L}|\omega|)}{\cosh(\gamma\omega)}e^{-\gamma|\omega|},  \eea
 which can be written as
 
 \bea E_{\ket{\frac{1}{2}}_{A_2}}-E_{\ket{-\frac{1}{2}}_{A_2}}=m_R-m_L. \eea
 
 Hence the ground state for odd number of sites is $\ket{\pm\frac{1}{2}}_{A_2}$ depending on the values of $h_L,h_R$.

\subsubsection{$A_2$: Even number of sites}
 
 The Bethe equations corresponding to all spin up reference state have two boundary string solutions $\lambda_{bs R}=\pi\pm i\gamma(1-\tilde{\epsilon}_{R})$, $\lambda_{bs L'}=\pi\pm i\gamma(1+\tilde{\epsilon}_{L})$. Adding $\lambda_{bs R}$ to the state $\ket{\frac{1}{2}}_{A_2}$  leads to the state with following root distribution

\bea\label{denbound1a2} \tilde{\rho}_{\ket{0}_{\beta A_2}} (\omega)=\tilde{\rho}_{\ket{\frac{1}{2}}_{A_2}}(\omega)+\Delta\tilde{\rho}_{R} (\omega), \eea
where $\Delta\tilde{\rho}_{R}$ is given by \ref{denbs} with $\beta=R$. The spin of the state containing this boundary string can be calculated using $S^z=\frac{N}{2}-M$, where
\bea \label{nroots22}M=1+\int_{-\pi}^{\pi}\rho_{\ket{0}_{R A_2}}(\lambda) d\lambda. \eea

We obtain $S^z_{\ket{0}_{R A_2}}=0$. The energy of the boundary string is given by \ref{bsenapp}, which is $-m_R$. 

\smallskip

 Adding the boundary string $\lambda_{bs L'}$ to the state $\ket{\frac{1}{2}}_{A_2}$, we obtain 
 
 \bea\label{denbound1a2} \tilde{\rho}_{\ket{0}_{L' A_2}} (\omega)=\tilde{\rho}_{\ket{\frac{1}{2}}_{A_2}}(\omega)+\Delta\tilde{\rho}_{L'} (\omega), \eea

where 

\bea\Delta\tilde{\rho}_{L'} (\omega)=-\frac{1}{4\pi}(-1)^{\omega}\frac{e^{-\gamma(3+\tilde{\epsilon}_{L})|\omega|}+e^{-\gamma(1-\tilde{\epsilon}_{L})|\omega|}}{1+e^{-2\gamma|\omega|}}.
\eea

The spin of the state containing this boundary string can be calculated using $S^z=\frac{N}{2}-M$, where
\bea \label{nroots23}M=1+\int_{-\pi}^{\pi}\rho_{\ket{0}_{L' A_2}}(\lambda) d\lambda. \eea

We obtain $S^z_{\ket{0}_{L' A_2}}=0$. The energy of the boundary string $\lambda_{bs L'}$ is given by

\bea E_{\lambda_{bs \beta '}}=-\frac{2\sinh^2\gamma}{\cosh\gamma+\cosh\gamma(1+\tilde{\epsilon}_{\beta})}+\sinh\gamma\sum_{\omega}(-1)^{\omega}e^{-2\gamma|\omega|}\frac{\cosh(\gamma(1+\tilde{\epsilon}_{\beta})\omega)}{\cosh(\gamma\omega)}=m_{\beta },\eea

 with $\beta=L$. The energy difference between the states $\ket{0}_{L' A_2}$ and $\ket{0}_{R A_2}$ can be calculated similar to the previous section, we obtain  
 
 \bea E_{\ket{0}_{L' A_2}}-E_{\ket{0}_{R A_2}} =  m_L+m_R. \eea

 Hence the ground state for even number of sites is $\ket{0}_{R A_2}$.

\subsubsection{$A_3$ and $A_4$ sub-phases}

In constructing a state in the phase $A_3$ or $A_4$, we can use the construction of the respective state in the phase $A_1$ or $A_2$ respectively, and use the following transformation:

\bea\label{z2bethe}
 |\uparrow\uparrow...\uparrow\rangle \leftrightarrow |\downarrow\downarrow...\downarrow\rangle, \hspace{5mm} h_L\rightarrow -h_L, \hspace{4mm}h_R\rightarrow h_R, \eea

where the all spin up and all spin down reference states are interchanged and the boundary magnetic fields change sign.

\subsection{Summary of Bethe Solution}
In this section we summarize the construction of the Bethe solution obtained above
\subsubsection{Odd number of sites}
\label{sec:Aoddnumber}
\paragraph{The $A_1$ and $A_3$ sub-phases.}
In  these cases  both  boundary magnetic fields point towards the same direction: along  the positive  $z$ axis for  the  $A_1$ sub-phase and negative $z$ axis for the  $A_3$ sub-phase. Both cases are related by the isometry (\ref{z2}). Qualitatively speaking, in the sub-phases $A_{1,3}$ and for $N$ odd, the boundary magnetic fields are not frustrating in the sense that in the Ising  limit of (\ref{hamiltonian})  the ground-state would exhibit perfect antiferromagnetic order.

1) In the $A_1$ phase we find that the ground-state is unique and has a total spin $S^z=-\frac{1}{2}$. It is constructed by starting with all spin down reference state and contains $\frac{N-1}{2}$ real roots. This state is labelled by $\ket{-\frac{1}{2}}_{A1}$.

2) In the $A_1$ phase, there exists an excited state with total spin $S^z=+\frac{1}{2}$, which does not contain any spinons. It is constructed by starting with all spin up reference state and contains $\frac{N-1}{2}$ real roots. This state is labelled by $\ket{+\frac{1}{2}}_{A1}$.

3)The energy difference between these two states in the $A_1$ phase is 

\bea E_{\ket{\frac{1}{2}}_{A_1}}-E_{\ket{-\frac{1}{2}}_{A_1}}=m_L+m_R.\eea

4) All the states in the $A_3$ phase can be obtained by using the symmetry \ref{z2bethe} described above.

\vspace{10mm}

\paragraph{The $A_2$ and $A_4$ sub-phases.}

In these cases the boundary fields are frustrating for $N$ odd  in the sense discussed above.

1) In the $A_2$ phase, for $|h_L|<|h_R|$, the ground state has total spin $S^z=-\frac{1}{2}$. It is constructed by starting with all spin down reference state and contains $\frac{N-1}{2}$ real roots. This state is labelled as $\ket{-\frac{1}{2}}_{A2}$.

2) In the $A_2$ phase, there exists an excited state with total spin $S^z=+\frac{1}{2}$, which does not contain any spinons. It is constructed by starting with all spin up reference state and contains $\frac{N-1}{2}$ real roots. This state is labelled by $\ket{+\frac{1}{2}}_{A2}$.

3)The energy difference between these two states in the $A_2$ phase is 

\bea E_{\ket{\frac{1}{2}}_{A_2}}-E_{\ket{-\frac{1}{2}}_{A_2}}=m_R-m_L. \label{diffena2odd}\eea

4) From the above expression, one can infer that in the $A_2$ phase, for $|h_L|>|h_R|$, the state $\ket{+\frac{1}{2}}_{A2}$ is the ground state and the state $\ket{-\frac{1}{2}}_{A2}$ is an excited state.

5) Using the symmetry \ref{z2}, we can obtain all the states in the sub-phase $A_4$ from the states in the sub-phase $A_2$.

\vspace{10mm}

\paragraph{Phase diagram}
Having obtained the solution in the region where $|h_{L,R}|<h_{c1}$ for odd number of sites chain, we can now obtain the phase diagram shown in the main text \ref{pdodd}. The region which corresponds to $h_R+h_L >0$, can be divided into three regions: (a) $h_R,h_L>0$  (b) $h_R>0, h_L<0, |h_R|>|h_L|$ (c)  $h_R<0, h_L>0, |h_R|<|h_L|$.

1) The region (a) is just the phase $A_1$ described previously. The ground state and the first excited states are $\ket{-\frac{1}{2}}$ and $\ket{\frac{1}{2}}$ respectively. 

2) The region (b) is contained within the phase $A_2$. From the energy difference between the states $\ket{-\frac{1}{2}}$ and $\ket{\frac{1}{2}}$ \ref{diffena2odd}, we can infer that the ground state and the first excited states are $\ket{-\frac{1}{2}}$ and $\ket{\frac{1}{2}}$ respectively.

3) The region (c) is contained within the phase $A_4$. The energy difference between the states 
$\ket{-\frac{1}{2}}$ and $\ket{\frac{1}{2}}$ can be obtained by using \ref{z2}, \ref{diffena2odd}:

\bea E_{\ket{\frac{1}{2}}_{A_2}}-E_{\ket{-\frac{1}{2}}_{A_2}}=-m_R+m_L \label{diffena4odd}\eea

Hence, the ground state and the first excited states are again $\ket{-\frac{1}{2}}$ and $\ket{\frac{1}{2}}$ respectively.

4) Using \ref{z2}, one can obtain all the states on the other side of the separatrix $h_R+h_L=0$, which corresponds to $h_L+h_R<0$. We find that the ground state and the first excited states in this region are $\ket{\frac{1}{2}}$ and $\ket{-\frac{1}{2}}$ respectively.

5) On the separatrix, using \ref{diffena2odd}, \ref{diffena4odd}, we find that the two states $\ket{-\frac{1}{2}}$ and $\ket{\frac{1}{2}}$ are degenerate.

\subsubsection{Even number of sites}
\label{sec:Aevennumber}
\paragraph{The $A_1$ and $A_3$ sub-phases.}

1)In the $A_1$ phase, for $h_R<h_L$, the ground state has total spin $S^z=0$. It is constructed by starting with all spin up reference state and contains $\frac{N-2}{2}$ real roots and the boundary string corresponding to the left edge $\lambda_{bs L}=\pi+\pm i\gamma(1-\tilde{\epsilon}_{L})$. This state is represented by $\ket{0}_{L,A_1}$. 

2) In the $A_1$ phase, for $h_R<h_L$, there exists an excited state which does not contain any spinons. This state has total spin $S^z=0$ and is constructed by starting with all spin up reference state and contains 
$\frac{N-2}{2}$ real roots and the boundary string corresponding to the right edge $\lambda_{bs R}=\pi+\pm i\gamma(1-\tilde{\epsilon}_{R})$. This state is represented by $\ket{0}_R$.

3) The energy difference between these two states is

\bea E_{\ket{0}_{L,A_1}}-E_{\ket{0}_{R,A_1}}=m_R-m_L. \label{diffena1even}\eea

4) One can infer from the above expression that, for $h_R>h_L$, the state $\ket{0}_{R,A_1}$ is the ground state and the state $\ket{0}_{L,A_1}$ is the excited state which does not contain any spinons.

5) Using the symmetry \ref{z2}, we can obtain all the states in the sub-phase $A_3$ from the states in the sub-phase $A_1$.

\paragraph{The $A_2$ and $A_4$ sub-phases.}

1) In the sub-phase $A_2$, the ground state has total spin $S^z=0$. It is constructed by starting with all spin up reference state and contains $\frac{N-2}{2}$ real roots and the boundary string corresponding to the right edge $\lambda_{bs R}=\pi+\pm i\gamma(1-\tilde{\epsilon}_{R})$. This state is represented by $\ket{0}_{R,A_2}$. 

2) In the sub-phase $A_2$, there exists an excited state which does not contain any spinons, and has total spin $S^z=0$. It is constructed by starting with all spin up reference state and contains $\frac{N-2}{2}$ real roots and the boundary string corresponding to the left edge $\lambda_{bs L'}=\pi+\pm i\gamma(1+\tilde{\epsilon}_{L})$. This state is represented by $\ket{0}_{L',A_2}$. 

3) The energy difference between these two states is

\bea E_{\ket{0}_{L' A_2}}-E_{\ket{0}_{R A_2}} =  m_L+m_R. \label{diffena2even}\eea

4) Using the symmetry \ref{z2}, we can obtain all the states in the sub-phase $A_4$ from the states in the sub-phase $A_2$.

\vspace{10mm}

\paragraph{Phase diagram}
Having obtained the solution in the region where $|h_{L,R}|<h_{c1}$ for even number of sites chain, we can now obtain the phase diagram for even number of sites \ref{pdeven}. The region which corresponds to $h_R>h_L$, can be divided into three regions: (d) $h_R>h_L>0$  (e) $h_R>0, h_L<0$ (f)  $h_R<0, h_L<0, |h_R|<|h_L|$.

1) The region (e) is just the phase $A_2$ described previously. The ground state and the first excited states have total spin $S^z=0$ containing boundary strings $\lambda_{bsR}$, $\lambda_{bsL'}$ respectively as discussed above. 

2) The region (d) is contained within the phase $A_1$. From the energy difference between the two states \ref{diffena1even}, we can infer that the ground state and the first excited states have total spin $S^z=0$ and contain the boundary strings $\lambda_{bsR}$, $\lambda_{bsL}$ respectively as described above.

3) The region (f) is contained within the phase $A_3$. The ground state and the first excited states and their energies can be obtained by using \ref{z2}, \ref{diffena1even}, and we find that the ground state and the first excited states are again same as in the region (e) described above.

4) Using \ref{z2}, one can obtain all the states on the other side of the separatrix $h_R=h_L$, which corresponds to $h_L>h_R$. We find that the ground state and the first excited states in this region are respectively the first excited state and the ground state corresponding to the region $h_L<h_R$ discussed above.

5) On the separatrix, using \ref{diffena1even}, \ref{diffena2even}, we find that these two states are degenerate.

\end{widetext}

\end{document}